\documentclass[preprint]{aastex}

\usepackage[latin1]{inputenc}                   
\usepackage{graphicx,color,textcomp}
\usepackage{natbib}

\begin{document}

\title{Gravitational lensing evidence against extended dark matter halos}

\author{Pierre Magain and Virginie Chantry}
\affil{Institut d'Astrophysique et de Géophysique, Université de Liège, Bat. B5C, Allée du 6 Août, 17, 4000 Sart Tilman, Liège 1, Belgium}

\email{Pierre.Magain@ulg.ac.be}

\begin{abstract}
It is generally thought that galaxies are embedded in dark matter halos extending well beyond their luminous matter.  The existence of these galactic halos is mainly derived from the larger than expected velocities of stars and gas in the outskirts of spiral galaxies.  Much less is known about dark matter around early-type (elliptical or lenticular) galaxies.  We use gravitational lensing to derive the masses of early-type galaxies deflecting light of background quasars.  This provides a robust measurement of the total mass within the Einstein ring, a circle whose diameter is comparable to the separation of the different quasar images.  We find that the mass-to-light ratio of the lensing galaxies does not depend on radius, from inner galactic regions out to several half-light radii.  Moreover, its value does not exceed the value predicted by stellar population models by more than a factor two, which may be explained by baryonic dark matter alone, without any need for exotic matter.  Our results thus suggest that, if dark matter is present in early-type galaxies, its amount does not exceed the amount of luminous matter and its density follows that of luminous matter, in sharp contrast to what is found from rotation curves of spiral galaxies.
\end{abstract}


\maketitle

\section{Introduction}

According to general relativity, light is deflected by massive objects.  This gives rise to the gravitational lensing phenomenon, where several images of a source located behind a mass concentration can be observed.  The deflecting object is called a gravitational lens.  In the present study, we consider systems in which the lens is a single galaxy and the background source a quasar, i.e. the bright nucleus of a distant galaxy.

If the source, lens and observer are perfectly aligned and if the lens is symmetrical around the line-of-sight, the image of the source is a ring (the Einstein ring) of angular radius
\begin{equation}
 \theta_{E} = \sqrt{\frac{4GM_E}{c^2}\frac{d_{LS}}{d_{OL}d_{OS}}}
\end{equation}
where $G$ is the gravitational constant, $M_E$ the mass of the lens inside the Einstein ring and $c$ the speed of light.  $d_{OL}$, $d_{OS}$ and $d_{LS}$ are the angular diameter distances from the observer to the lens, from the observer to the source and from the lens to the source.  Given a cosmological model, these distances can be computed from the redshifts of the lens and source spectra.  All distances being inversely proportional to the Hubble constant $H_0$, which measures the rate of expansion of the Universe, the uncertainty on the mass for a given Einstein radius will reflect the present uncertainty on the Hubble constant, i.e. less than 10\% \citep[see, e.g.,][ for a determination with a claimed 3\% precision]{Riess2011}.  The size of the Einstein ring thus provides a direct and robust measurement of the mass of the lens galaxy up to the angular distance $\theta_{E}$.

If the source and lens are slightly misaligned, one observes four images of the quasar, often associated to arcs which are distorted images of the quasar host galaxy.  Such a configuration allows to strongly constrain the size of the Einstein ring and, thus, the lens mass inside it.  If the misalignment is somewhat larger, only two images of the source are observed.  In this case, the constraints are somewhat weaker, but the mass can still be determined with an accuracy of $\sim$30\%, the main uncertainty coming from the shape of the mass distribution in the lens.

\section{Method}

We consider a sample of 25 gravitational lenses with high precision analysis \citep{Cosmograil8,Cosmograil10} on the basis of near-infrared H-band images ($1.63 \ \mu m$) obtained with the Hubble Space Telescope (HST). These objects were initially chosen because they had measured time delays or because they were part of the COSMOGRAIL observing campaign. The images were processed with the iterative version of the MCS deconvolution algorithm \citep{MCS98,Chantry2007} in order to get accurate angular positions (i.e. from 1 to 2.5 mas) for the quasar images as well as to determine the shape, position and luminosity of the lens galaxy. These lenses have then been modeled with the LENSMODEL software of Keeton \citep{Keeton2001} to obtain the Einstein radius and thus the mass inside it.  The full details of the image processing and lens modelling are provided in \cite{Cosmograil8} and \cite{Cosmograil10}.

From this sample, we exclude the objects for which the mass-to-light ratio of the lens might not be obtained with high accuracy. These include systems for which the redshift of the lens or of the source is not securely known, those with several galaxies of similar luminosity acting as lenses and those for which a clear separation of the light from the lens galaxy and from the source images could not be guaranteed.  This leaves us with a sample of 15 lenses, 7 doubles and 8 quadruples.  Figure 1 shows the deconvolved HST images of these lenses.

The flux inside the Einstein ring of each lens galaxy is determined from the deconvolved HST images, after careful removal of the quasar images as well as associated arcs, when present.  These fluxes are corrected for the redshift of the lens using a k-correction based on elliptical galaxy template spectra \citep{Fioc1997}.  These k-corrections, which account for the shift of the lens spectrum with respect to the filter bandpass, are quite secure in the near-infrared, where the spectral energy distribution depends only weakly on the galaxy type.  The solar luminosity in the H-band \citep{Colina1996} is computed at the distance of each lens galaxy using the luminosity distance \citep{Wright2006} for a cosmological model with $H_0 = 70$ km/s/Mpc, $\Omega_m = 0.27$ and $\Omega_{\Lambda} = 0.73$.  The lens galaxy luminosity can then be computed in solar units, as well as an error bar taking into account the quality of the separation between source and lens features.

For sufficiently symmetric lenses, the half-light radius $R_{1/2}$ (i.e. the radius of the aperture containing one half of the light emitted by the galaxy) is determined by directly measuring the light flux $F$ in circles of varying radii $R$ and computing the slope of $\log{F}$ versus $R^{1/n}$, $n$ being chosen so that the slope remains as constant as possible with radius. In the remaining cases, it is determined by fitting Sérsic profiles \citep{Sersic1963} on the lens galaxies with the Galfit software \citep{Peng2002,Peng2010}.  Final values and error bars are presented in Table 1.

Most of the past determinations of the dark matter fraction in early-type galaxies \citep[e.g.][]{Koopmans2006,Bolton2008a} have been restricted to the inner galactic regions, i.e. $R_E \le R_{1/2}$).  However, in order to constitute a clear-cut test of the dark matter halo hypothesis, the mass determinations should extend to regions well outside the bulk of the visible galactic matter. This can be quantified by the radius of the Einstein ring $R_E$ in units of the half-light radius $R_{1/2}$. Our determinations range from $R_E = R_{1/2}$ to $R_E = 5 R_{1/2}$.  This can be compared to studies of rotation velocities. As an example, for the bright spiral galaxy NGC 3198, which is one of the best studied objects, the optical rotation curve \citep{Corradi1991} extends to 1.8 $R_{1/2}$ while the 21 cm hydrogen line measurements \citep{vanAlbada1985} reach 5.8 $R_{1/2}$. Our determinations thus go well beyond the optical observations and reach nearly as far as the best radio measurements in spiral galaxies.

\section{Results and discussion}

Our results are presented in Fig. \ref{ML}, which shows that the $M/L_H$ ratio is about 1.8, independent of $R_E/R_{1/2}$.  Such a constant M/L ratio indicates that the mass distribution closely follows the light distribution and does not show evidence for an extended dark matter halo. A least squares linear regression taking into account errors in both coordinates gives a slope of $0.12 \pm 0.13$ (one sigma error bar), which is totally compatible with zero.

Our $M/L_H$ ratios can be compared to theoretical predictions from stellar population models \citep{Maraston2005}.  Theoretical values lie between 0.66 and 1.69, depending on the initial mass function and on the age of the stellar population, here taken to lie between 5 and 10 billion years, which is a reasonable range for elliptical galaxies with redshifts from 0.2 to 1.0.  

The theoretical $M/L_H$ ratios, which are for stellar populations only, are generally lower than the ones deduced from gravitational lensing.  This can be put in a cosmological context by comparing the estimate of baryonic density in the Universe deduced from the results of Big Bang nucleosynthesis \citep{Schramm}, i.e.\   4\% of the critical density, with the density obtained from an inventory of currently detected matter \citep{Fukugita,Cen}, amounting to about 2\% of the critical density.  This comparison indicates that about one half of the baryonic matter remains undetected.  Our results are consistent with this, and suggest that the hidden baryonic matter might be found in galaxies, with a distribution comparable with that of luminous matter.

The main goal of the present paper is to test the hypothesis that early-type lens galaxies are embedded in extended dark matter halos, as are generally found around spiral galaxies.  The constancy of the $M/L_H$ ratio as a function of radius provides evidence against that hypothesis.  For further illustration, we have computed the mass expected in the form of a typical dark matter halo in the following way.  We assume that the visible mass is distributed in order to emit light according to a de Vaucouleurs profile, i.e. a Sérsic profile with n=4 \citep[]{Mellier1987}, which is typical for elliptical galaxies.  We assume spherical symmetry and add a dark matter halo with a total mass 4 times the galactic mass within 5.8 $R_{1/2}$, as derived from radio observations of NGC 3198 \citep{vanAlbada1985}.  This gives a nearly flat velocity curve, as usually observed.  The total $M/L$ ratio is then computed as a function of $R/R_{1/2}$ and compared to our determinations.

Fig.\  2 shows that our results do not agree with the hypothesis of extended dark matter halos, as the ones required to produce approximately flat velocity curves in spiral galaxies.  It should be reminded that, while the main evidence for dark matter halos comes from the rotation curves of spiral galaxies, the lens galaxies analyzed here are ellipticals, for which it is much more difficult to derive velocity diagnostics.  Some controversy has arisen about the presence of dark matter halos around elliptical galaxies \citep{Romanowsky2003,Dekel2005}.  However, massive ellipticals are generally considered as the result of fusion of spiral galaxies.  It is thus hard to understand how dark matter halos would be present around spirals and absent after their fusion.

Some additional evidence is provided by a number of recent gravitational lensing studies which find a good correlation of the ellipticity and position angle of the total mass with those of luminous matter \citep{Cosmograil10,Koopmans2006,Gavazzi2012}.  Adding our result, we can conclude that the ellipticity, position angle and length scale of the total mass are strongly correlated with those of luminous matter.  This suggests that the total mass distribution in early-type galaxies closely follows the light distribution and sheds doubts on the existence of extended galactic halos made of exotic, non-baryonic particles.

We do not ignore the many successes of the dark matter hypothesis, which helps solving a number of problems in cosmology.  However, it has also experienced a number of failures \citep[see, e.g.,][]{Kroupa2012} which, together with the present results, suggest that alternative explanations might be worth investigating.

\begin{figure*}
\label{Images}
\includegraphics[width=0.9\textwidth]{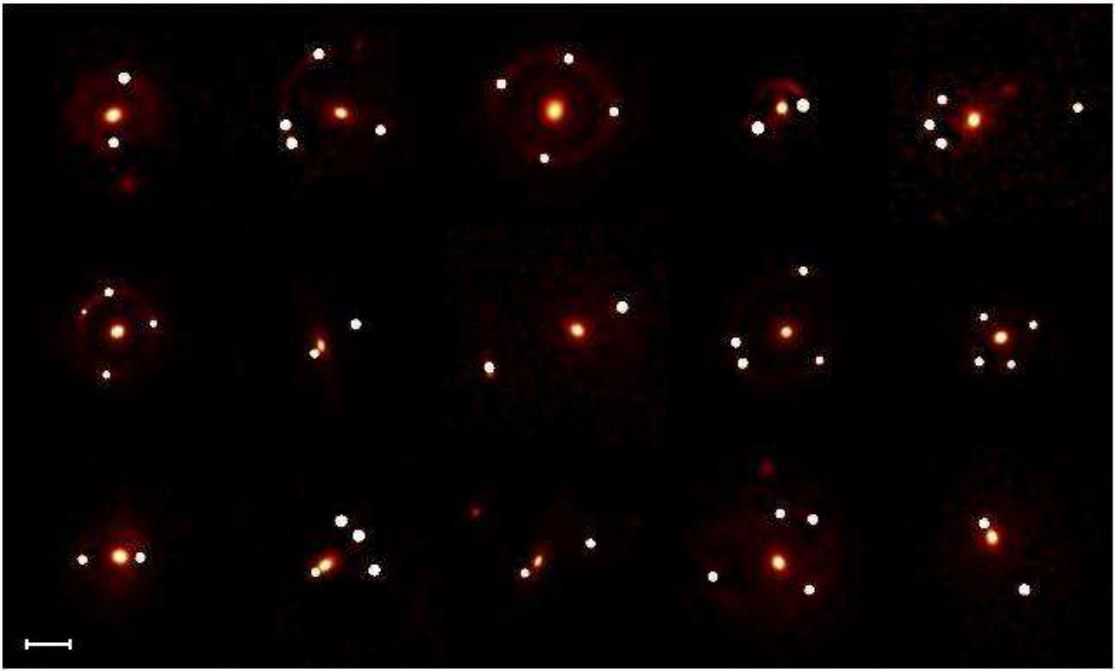}
\caption{Deconvolved HST images of the 15 gravitational lens systems, shown on the same angular scales and orientation (North up and East to the left).  Intensities are scaled on the centre of the lens galaxy.  Images and spectra indicate that most of the lens galaxies are early-type (i.e. elliptical or lenticular galaxies).  The lens systems are ordered as in Table 1.  The images of MG0414+0534 and SDSS0924+0219, which had not been previously deconvolved in \cite{Cosmograil8} and \cite{Cosmograil10}, have been processed the same way for the present work.  The horizontal bar corresponds to 1 arcsecond.}
\end{figure*}

\begin{figure*}
\label{ML}
\includegraphics[width=0.8\textwidth]{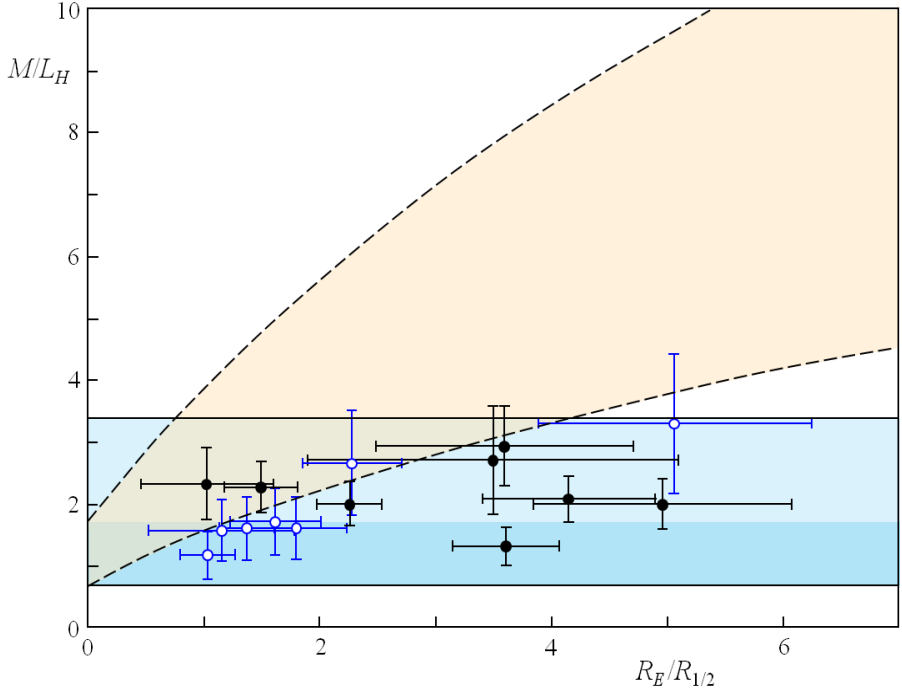}
\caption{Measured near-infrared mass-to-light ratios within the Einstein rings of the lens galaxies compared to expectations with and without a dark matter halo.  The blue open circles correspond to the double lenses while the full black circles are for the quadruple lenses, which provide the best constraints on masses.  The darker blue rectangle corresponds to stellar population model predictions.  The lighter blue rectangle corresponds to the addition of baryonic dark matter with the same distribution as the visible matter.  The orange area corresponds to models with a dark matter halo.   Data points in the brownish intersecting area are compatible with both hypotheses, as expected for the inner regions of galaxies.  The dark matter model is an isothermal sphere with core radius of 0.5 $R_{1/2}$ added to baryonic matter with $M/L_H$ in the adopted range for stellar populations (an NFW model \citep{NFW1996} with a scale radius of 2.0 $R_{1/2}$ gives very similar results).  The halo models are calibrated to give a dark matter mass equals to 4 times the visible mass within 5.8 $R_{1/2}$.}
\end{figure*}

\begin{table*}
\label{LensData}
\centering 
\begin{tabular}{c||ccccccc}
Object & N & $z_{lens}$ &  $z_{src}$ & $R_{1/2}$ & $R_E$ & $L_E$ & $M_E$ \\ 
\hline
HE0047--1756 & 2 & 0.41 & 1.66 & 3.00$\pm$0.30 & 4.11 & 7.26$\pm$0.73 & 11.70 \\
MG0414+0534 & 4 & 0.96 & 2.64 & 2.50$\pm$0.45 & 8.88 & 27.20$\pm$2.70 & 52.57 \\
HE0435--1223 & 4 & 0.45 & 1.69 & 6.80$\pm$3.80 & 6.97 & 14.30$\pm$2.80 & 33.43 \\
SBS0909+523 & 2 & 0.83 & 1.38 & 0.85$\pm$0.15 & 4.30 & 5.92$\pm$0.95 & 19.57 \\
RXJ0911+0551 & 4 & 0.77 & 2.80 & 4.60$\pm$0.90 & 6.89 & 12.35$\pm$1.24 & 28.14 \\
SDSS0924+0219 & 4 & 0.39 & 1.52 & 0.94$\pm$0.20 & 4.66 & 7.76$\pm$1.00 & 15.57\\
FBQ0951+2635 & 2 & 0.26 & 1.24 & 1.90$\pm$1.00 & 2.19 & 2.59$\pm$0.26 & 4.09\\
HE1104--1805 & 2 & 0.73 & 2.32 & 4.46$\pm$0.50 & 10.18 & 24.50$\pm$2.50 & 65.43\\
PG1115+080 & 4 & 0.31 & 1.72 & 1.97$\pm$0.20 & 4.45 & 7.28$\pm$0.73 & 14.57\\
SDSS1138+0314 & 4 & 0.45 & 2.44 & 1.07$\pm$0.11 & 3.86 & 6.85$\pm$1.20 & 9.10\\
SDSS1226--0006 & 2 & 0.52 & 1.12 & 3.44$\pm$0.60 & 3.56 & 9.32$\pm$0.93 & 11.01\\
B1422+231 & 4 & 0.34 & 3.62 & 1.00$\pm$0.30 & 3.59 & 3.17$\pm$0.50 & 9.33\\
SBS1520+530 & 2 & 0.76 & 1.86 & 3.10$\pm$0.60 & 5.58 & 14.00$\pm$1.40 & 22.57\\
WFI2033--4723 & 4 & 0.66 & 1.66 & 3.75$\pm$1.70 & 13.10 & 12.00$\pm$3.50 & 32.57 \\
HE2149--2745 & 2 & 0.60 & 2.03 & 3.60$\pm$0.70 & 5.82 & 12.60$\pm$1.50 & 21.71 \\
\hline
\end{tabular}
\caption{Gravitational lens data. The columns list (1) the name of the object, (2) the number of images, (3) the lens galaxy redshift, (4) the source redshift, (5) the H-band half-light radius (in kpc), (6) the Einstein radius (in kpc), (7) the H-band luminosity inside the Einstein radius (in $10^{10}$ solar luminosities) and (8) the mass inside the Einstein radius (in $10^{10}$ solar masses).  The adopted (conservative) uncertainties on $R_E$ are 15\% for the doubles and 7.5\% for the quadruples; the corresponding uncertainties on $M_E$ are 30\% and 15\%, respectively.}
\end{table*}

\acknowledgments

We wish to thank Dominique Sluse for the computation of the lens masses and for critical discussion.  This work has been supported by ESA and the Belgian Science Policy Office under the PRODEX program PEA 90312.


\bibliographystyle{apj}

\end{document}